\documentclass[aps,prl,twocolumn,showpacs]{revtex4}
\usepackage{amsmath,amssymb}
\usepackage{float}
\usepackage[pdftex]{graphicx}
\usepackage{subfigure}
\newcommand \beq{\begin{eqnarray}}
\newcommand \eeq{\end{eqnarray}}

\begin{document}

\title{Quasi-Nambu-Goldstone Modes 
in Bose-Einstein Condensates}

\author{Shun Uchino$^{1}$, Michikazu Kobayashi$^{2}$,
Muneto Nitta$^{3}$, and Masahito Ueda$^{1,4}$}

\affiliation{$^{1}$Department of Physics, The University of Tokyo,
7-3-1 Hongo, Tokyo 113-0033, Japan\\
$^2$Department of Basic Science, The University of Tokyo, 3-8-1 Komaba, 
Tokyo 153-8902, Japan\\
$^3$Department of Physics, and Research and Education Center for Natural
Sciences, Keio University, 4-1-1 Hiyoshi, Kanagawa 223-8521, Japan\\
$^{4}$ERATO Macroscopic Quantum Project, JST, Tokyo 113-8656, Japan
}

\date{\today}

\begin{abstract}
We show that quasi-Nambu-Goldstone (NG) modes,
which play prominent roles in high energy physics but have been
elusive experimentally,
can be realized with atomic Bose-Einstein condensates.
The quasi-NG modes emerge when the symmetry of a ground state
is larger than that of the Hamiltonian.
When they appear, the conventional vacuum manifold
should be enlarged.
Consequently topological defects that are stable within 
the conventional vacuum manifold become unstable and decay by emitting 
the quasi-NG modes. Contrary to conventional wisdom,
however, we show that the topological defects
are stabilized by quantum fluctuations that
make the quasi-NG modes massive, thereby suppressing their emission.
\end{abstract}


\maketitle

\emph{Introduction.}--- 
Spontaneous symmetry breaking
generally
yields Nambu-Goldstone (NG) modes, which
play a crucial role 
in determining low-energy behaviors of 
various systems from condensed matter to high energy physics.
Although the NG theorem guarantees that
NG modes do not acquire mass at any order of quantum corrections,
we sometimes encounter situations in which soft modes appear 
which are massless
in the zeroth order (tree approximation) but become massive due to
quantum corrections.
They are called quasi-NG modes \cite{note1}
and were
introduced  in the context of gauge theories with symmetry breaking
by Weinberg \cite{weinberg}.
It was shown that
such modes emerge if the symmetry of an
effective potential of the zeroth order
is higher than that of gauge symmetry, and 
the idea  was invoked to account for the emergence of
low-mass particles. On the other hand,
Georgi and Pais demonstrated that
quasi-NG modes also occur in cases in which the symmetry of the ground state 
is higher than that of the Hamiltonian \cite{georgi}.

Later on, quasi-NG modes turned out to play prominent roles in
technicolor \cite{weinberg2} and supersymmetry \cite{kugo}, 
both of which were proposed as 
candidates beyond the standard model for unification of fundamental forces.
The quasi-NG modes of the Weinberg type
\cite{weinberg}
become an important ingredient 
in the physics of technicolor,
which is a model
to avoid a hierarchy problem because it does not assume the Higgs
scalar field
as an elementary particle.
In this model,
the quasi-NG mode is related to  
the vacuum that is energetically selected 
from among a large degenerate family of vacua (vacuum alignment problem). 
On the other hand,
the quasi-NG modes of the Georgi-Pais type \cite{georgi}
are inevitable in theories with supersymmetry,
which is among the most powerful guiding principles in
contemporary elementary particle physics.
This is because the vacuum condition in supersymmetric
theories invariant under a group $G$ is always 
invariant under 
its complex extension $G^{\mathbb{C}}$.
This type of the quasi-NG mode
is also believed to appear in the weak-coupling limit of 
$A$-phases of $^{3}$He \cite{volovik} and spin-1 color superconductivity
\cite{pang}. 
Despite their importance as described above,
the direct experimental confirmation of the quasi-NG modes
has yet to be made.
\begin{table}
\caption{Vacuum manifold $M$ and the
enlarged vacuum manifold $\tilde{M}$ of
the uniaxial nematic $(\eta=n\pi/3)$, biaxial nematic $[\eta=(n+1/2)\pi/3]$, 
and dihedral-2 (other $\eta$) phases,
where the parameter $\eta$ characterizes the order parameter of the
nematic
phase [see Eq. \eqref{nematic-1}].}
\label{table}
\begin{center}
\begin{tabular}{l|c|c}
\hline
Phase & $M\cong G/H$ & $\tilde{M}$   \\ 
\hline
Uniaxial nematic & $\text{U(1)}\times\text{S}^2/\mathbb{Z}_2$  &   \\ 
Biaxial nematic 
& $[\text{U(1)}\times\text{SO(3)}]/\text{D}_4$ 
& $[\text{U(1)}\times\text{S}^4]/\mathbb{Z}_2$ \\
Dihedral-2 & $\text{U(1)}\times\text{SO(3)}/\text{D}_2$ &  \\
\hline
\end{tabular}
\end{center}
\end{table}

\begin{figure}
\includegraphics[width=3in]{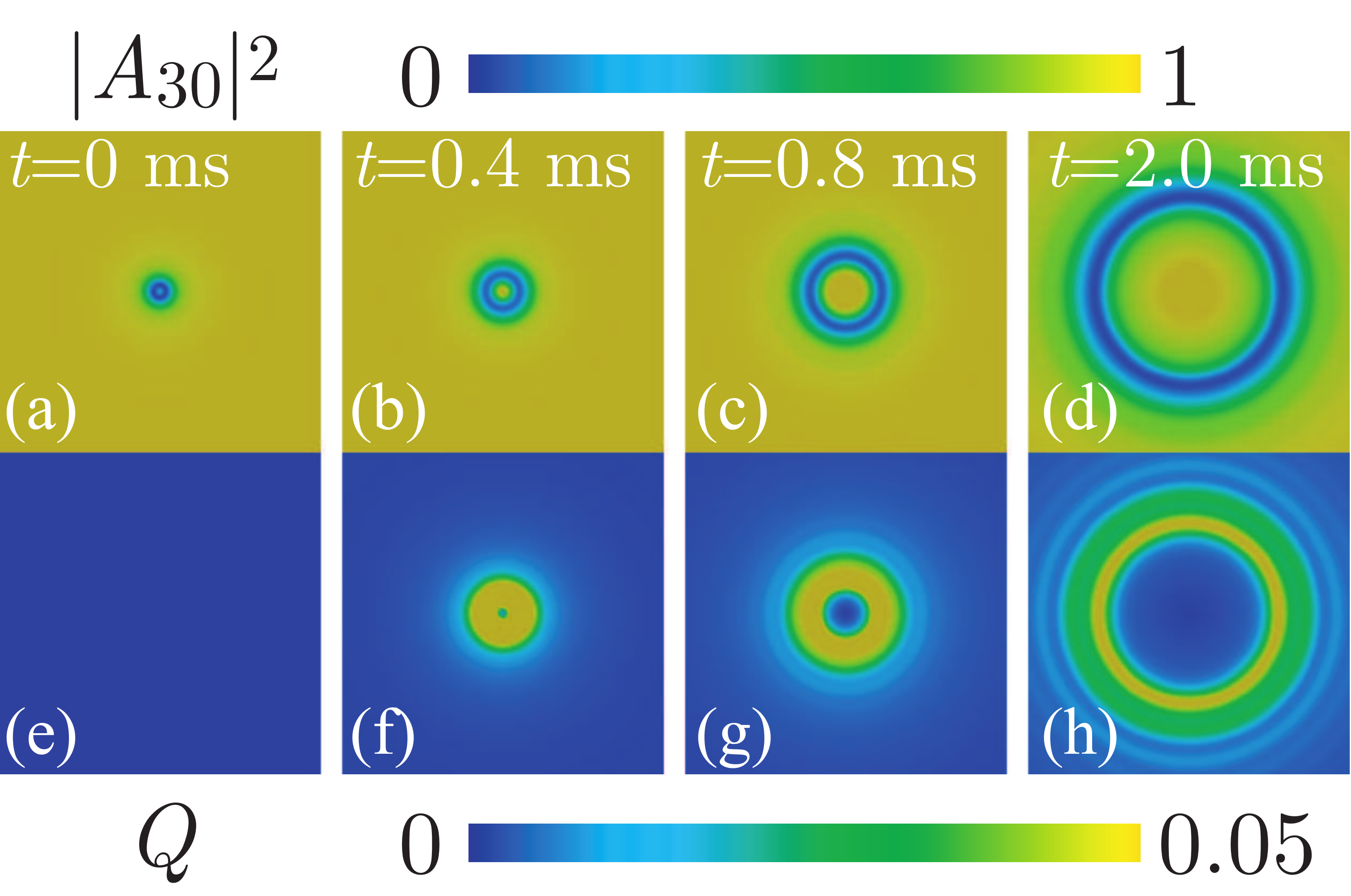}
\caption{(color online).
Decay dynamics of a $\mathbb{Z}_2$ vortex
that is stable under 
$M\cong \text{U(1)}\times\text{S}^2/\mathbb{Z}_2$ in the uniaxial
nematic phase.
(a)-(d) Time evolution of the absolute square of the spin-singlet trio
amplitude $|A_{30}|^2$ defined in Eq. \eqref{trio}. 
(e)-(h) Time evolution of the quasi-NG charge
$Q\equiv \langle F_{35}\cos\theta+F_{13}\sin\theta\rangle$, where
$\theta$ is the polar coordinate on a plane and $F_{13}$ and $F_{35}$ are
introduced below Eq. \eqref{op-bn-fn}.
The quantum fluctuations that stabilize the vortex state
are switched off at $t=0$. Then the vortex decays in time as shown in (b)-(d).
This decay is accompanied by the emission of quasi-NG modes as shown in
 (f)-(h).
The numerical simulations are performed by using a modified
 Gross-Pitaevskii 
analysis
based on a local density approximation \cite{fabrocini}.
In the simulations, we choose numerical parameters for
typical experimental
situations for spin-2 $^{87}$Rb condensates 
with $n=1.5\times 10^{14}\text{cm}^{-3}$.}
\label{fig1}
\end{figure}

In this Letter, we point out that a spinor Bose-Einstein condensate  
is an ideal system to study the
physics of quasi-NG modes.
This system
has recently been a subject of active research because of the 
great experimental manipulability and
well-established microscopic Hamiltonians.
We show that quasi-NG modes appear in a spin-2 nematic
phase, which may be realized by using 
$^{87}$Rb and
a $d$-wave superfluid. 
In the nematic condensate, three phases, 
each of which has a different symmetry, 
are energetically degenerate to the zeroth order \cite{barnett}
as listed in Table \ref{table}.
We point out that this corresponds to the vacuum alignment
problem in particle physics, and
to fully understand this, we must consider the degrees of freedom of
quasi-NG modes due to the $\text{U(1)}\times\text{SO(5)}$ symmetry to
the zeroth order.
Considering this symmetry,
we show that the vacuum (order parameter) manifold is
enlarged to   
$\tilde{M}\cong [\text{U(1)}\times\text{S}^4]/\mathbb{Z}_2$.
Then, the vacuum alignment problem can be understood
from the fact that each vacuum manifold of the NG modes $M\cong G/H$ 
is a submanifold of $M$.
We show that the vacuum alignment problem is solved because of 
the quasi-NG modes becoming massive and that one of
the vacuum manifolds is selected because the symmetry at the zeroth order
is broken by quantum corrections.
We find that a soliton that is unstable classically
becomes stabilized by quantum corrections that
suppress emissions of the quasi-NG modes as shown in Figs. 
\ref{fig1}(a) and (e).
Notably, as soon as the quantum fluctuations are switched off at $t=0$,
the soliton begins to decay since the original vortex core structure disappears 
as shown in
Figs. (b)-(d); this is accompanied by emissions of quasi-NG modes as shown in 
Figs. (f)-(h).

\emph{Symmetry of the ground state.}--- 
We start with a spin-2 condensate with mass
$M$ whose $G=\text{U(1)}\times\text{SO(3)}$ invariant
Hamiltonian is given by
\beq
H=\int d^3r\Big[ -\frac{\hbar^2}{2M}\psi^{\dagger}_m\Delta\psi_m
+\frac{c_0}{2}(\psi^{\dagger}_m\psi_m)^2\nonumber\\
+\frac{c_1}{2}(\psi^{\dagger}_m\mathbf{F}_{mn}\psi_n)^2
+\frac{c_2}{2}|\psi^{\dagger}_m\textit{T}\psi_m|^2\Big],
\label{total-hamiltonian}
\eeq
where $\psi_m$ $(m=2,1,...,-2)$ are field operators
with magnetic quantum number $m$, $F_i$ $(i=x,y,z)$ are spin matrices,
$\textit{T}$ denotes the time reversal operator 
[$T\psi_m=(-1)^m\psi_{-m}^{\dagger}$], 
and repeated indices 
are assumed to be summed over $2,1,...,-2$. 
The coupling constants are related to the $s$-wave scattering lengths 
$a_S$ in the total spin $S$ channel by
$c_0=4\pi\hbar^2(4a_2+3a_4)/7M$, $c_1=4\pi\hbar^2(a_4-a_2)/7M$, and
$c_2=4\pi\hbar^2(7a_0-10a_2+3a_4)/35M$.
By applying the Gross-Pitaevskii theory to the Hamiltonian 
\eqref{total-hamiltonian} under the assumption that 
the $\mathbf{k}=\mathbf{0}$ components are macroscopically occupied, 
we can show that 
the ground state of the nematic phase is realized for $c_2<0$ and
$c_2<4c_1$ \cite{koashi,ciobanu}. 
The order parameter of the nematic phase
is given by
\beq
\phi_m=
(\sin\eta/\sqrt{2},0,\cos\eta,0,\sin\eta/\sqrt{2})^T,
\label{nematic-1}
\eeq
where $\eta$ is an additional parameter independent of 
$G$ \cite{barnett}, and 
$\eta=n\pi/3$, $\eta=(n+1/2)\pi/3$, and
the other values correspond to the uniaxial nematic,
biaxial nematic, and dihedral-2 phases, respectively,  
where $n\in \mathbb{Z}$.
Since the state with $\eta$ is equivalent to that with $\eta+\pi/3$
up to U(1)$\times$SO(3), 
we restrict the domain of definition to
$0\le\eta <\pi/3$ 
unless otherwise stated. 

Next, we show that the zeroth-order solution of the nematic phase
has $\tilde{G}=$U(1)$\times$SO(5) symmetry.
As discussed in Ref. \cite{uchino}, 
the interactions proportional to $c_0$, $c_1$, and $c_2$ have
SU(5), SO(3), and SO(5) symmetries, respectively, 
in addition to the U(1) symmetry.
On the other hand, the zeroth-order solution for the nematic phase satisfies
$|\phi^{\dagger}_m
\textit{T}\phi_m|=1$
and $\phi^{\dagger}_m(\mathbf{F})_{mn}\phi_n=\mathbf{0}$.
In fact, the first relation
is sufficient to characterize the nematic phase, since it is proved that
$|\langle\psi^{\dagger}_m\textit{T}\psi_m\rangle|=1$ implies
$\langle\mathbf{F}\rangle=\mathbf{0}$ \cite{ciobanu}.
Within the zeroth order, the interaction term proportional to $c_2$ 
satisfies $|\phi^{\dagger}_m\textit{T}\phi_m|^2=1$,
and is invariant under a $\tilde{G}$ transformation, which implies
that $|\phi^{\dagger}_m\textit{T}\phi_m|=1$ is
preserved under the same transformation.
Therefore, we conclude that the solution of the nematic phase
has the $\tilde{G}$ symmetry even if $c_1\ne 0$.

\emph{Number of NG and quasi-NG modes.}---
We identify the enlarged vacuum manifold of the nematic
phase based on $\tilde{G}$ and 
discuss the number of NG and quasi-NG modes in each phase.
Since the vacuum manifold is independent of the location in the orbit,
we choose the configuration
$\bar{\phi}_m=(-i,0,0,0,i)^T/\sqrt{2}$,
which describes the five-dimensional representation of SO(3) in the
spherical
tensor basis
and therefore describes the SO(5) in the same basis. 
To determine the isotropy group, 
let us next transform from this representation of SO(5)
to the
fundamental representation of SO(5). 
By using a unitary matrix
\beq
U=\frac{1}{\sqrt{2}}
\begin{pmatrix}
	i & 0& 0 & 0 & -i \\
	0 & -i & 0 & -i & 0 \\
	0 & 0 & \sqrt{2} & 0 & 0 \\
        0 & 1 & 0 & -1 & 0 \\
        -1 & 0 & 0 & 0 & -1
\end{pmatrix},
\label{unitary}
\eeq 
we can achieve it for the state and generators as
\beq
\varphi_{k}\equiv U\bar{\phi}_{m}=(1,0,0,0,0)^T
\label{op-bn-fn}
\eeq
and $(\chi_{kl})_{ij}\equiv (UF_{kl}U^{\dagger})_{ij}
=-i(\delta_{ki}\delta_{lj}-\delta_{li}\delta_{kj})$,
respectively, where
$F_{kl}$ is the corresponding SO(5) generator in the spherical tensor
basis.
In the fundamental representation, elements of the SO(5) group are
expressed as $R_{kl}(t)=\exp (-i\chi_{kl}t)$ with real parameter $t$.
Then, \eqref{op-bn-fn} is invariant under
$R_{23}$, $R_{24}$, $R_{25}$,
$R_{34}$, $R_{35}$, and $R_{45}$, which
constitute an SO(4) as a subgroup of SO(5).
Therefore, the enlarged vacuum manifold except for discrete groups is
U(1)$\times$SO(5)/SO(4)$\cong$U(1)$\times$S$^4$.
To consider the effect of discrete symmetry,
let us utilize the degrees of freedom of
U(1)$\times$S$^4$. Then,
$\varphi_k$ can be transformed as follows:
\beq
\varphi'_k=e^{i\phi}(\cos\theta_1,\sin\theta_1\cos\theta_2,
\sin\theta_1\sin\theta_2\cos\theta_3,\nonumber\\
\sin\theta_1\sin\theta_2\sin\theta_3\cos\theta_4,
\sin\theta_1\sin\theta_2\sin\theta_3\sin\theta_4)^T,
\eeq
where $0\le\theta_1,\theta_2,\theta_3\le \pi$
and $0\le\theta_4\le 2\pi$.
To ensure that $\varphi'_k$ coincides with $\varphi_k$,
it is necessary and sufficient that $\theta_1=\phi=0$, or
$\theta_1=\phi=\pi$,
which is isomorphic to $\mathbb{Z}_2$.
The full enlarged vacuum manifold of the nematic phase is therefore given by
\beq
\tilde{M}\cong 
\frac{\text{U(1)}\times\text{S}^4}{\mathbb{Z}_2}
\cong \frac{\text{U(1)}\times\text{SO(5)}}{\mathbb{Z}_2\ltimes\text{SO(4)}}.
\label{opm}
\eeq
Here, $\ltimes$ implies that
the nontrivial element of $\mathbb{Z}_2$ does not commute with some elements 
of SO(4). 

Next, we discuss the number of NG and quasi-NG modes.
As analyzed in Ref. \cite{uchino2},
the number of the NG modes in the nematic phase is equal to
the dimension of $M$, dim($M$).
Meanwhile, in Ref. \cite{georgi},
it has been shown that the number of quasi-NG modes, $n$, is given by
\beq
n=\text{dim}(\tilde{M})-\text{dim}(M),
\label{number-png}
\eeq
where $\tilde{M}$ is the surface on which the effective potential
assumes its minimum value to the zeroth order.
This implies that $M$ is a submanifold of $\tilde{M}$ and
$n$ is the dimension of the complementary space of $M$ inside $\tilde{M}$.

In the nematic phase, we expect  
5 soft modes because  $\text{dim}(\tilde{M})=5$, which
is consistent with the result of the Bogoliubov theory
since it predicts an equal number of  massless modes
\cite{zhou,uchino2}.
While three and two of them are the NG and quasi-NG modes for the
uniaxial nematic phase 
because of dim($M$)=3,
four and one of them are the NG and quasi-NG modes 
for the dihedral-2 and biaxial nematic phases
because of dim($M$)=4. 
(See Table \ref{table1}.)
That is, the number of NG and quasi-NG modes changes in each phase,
with the total number unchanged.
\begin{table}[t]
\caption{Broken generators within \eqref{nematic-1} and $0\le\eta
 <\pi/3$, 
the number of
ordinary NG modes [U(1), $F_x$, $F_y$, and $F_z$], 
and that of quasi-NG
modes ($F_{13}$ and $F_{35}$) in each phase, where
$F_{x}=-F_{14}-F_{25}+\sqrt{3}F_{23}$,
$F_{y}=-F_{12}+F_{45}-\sqrt{3}F_{34}$, and
$F_{z}=2F_{15}+F_{24}$. }
\label{table1}
\begin{center}
\begin{tabular}{l|c|c|c}
\hline
Phase &   Broken generator & $N_{\text{NG}}$ &$N_{\text{quasi-NG}}$ \\ 
\hline
Uniaxial nematic &   
U(1)$,F_x,F_y,F_{13},F_{35}$ & 3 & 2 \\ 
Biaxial nematic & 
U(1)$,F_x,F_y,F_z,F_{35}$ & 4 & 1 \\
Dihedral-2 & 
U(1)$,F_x,F_y,F_z,F_{35}$& 4 & 1 \\
\hline
\end{tabular}
\end{center}
\end{table}

\emph{Quantum symmetry breaking.}---
Since the interactions proportional to $c_0$ and $c_2$
favor $\tilde{G}$ but the interaction proportional to
$c_1$ breaks $\tilde{G}$,
it is expected that the ground-state symmetry 
of the nematic phase at the zeroth order is
broken by quantum corrections, thereby making quasi-NG modes
massive.
In fact, the ground-state energy per particle at the 1-loop level
is evaluated as \cite{zhou,uchino2}
\beq
\Delta \epsilon&=&v(\eta)+\tilde{v},\\
v(\eta)&=&\omega
\textstyle\sum_{j=0}^2 
(1+X\cos(2\eta+2\pi j/3))^{5/2},
\label{qsb}
\eeq
where $\omega=8\sqrt{M^3}[n(2c_1-c_2)]^{5/2}/15n\pi^2\hbar^3$,
$X=-2c_1/(2c_1-c_2)$, $\tilde{v}$ describes an $\eta$-independent contribution,
and $v(\eta)$ is the $\tilde{G}$ symmetry breaking contribution, 
which favors the uniaxial nematic phase for $c_1\ge 0$
and the biaxial nematic phase for $c_1\le 0$.
In other words, 
the degeneracy at the zeroth order is lifted and one of the phases is
selected, depending on the sign of $c_1$.
This phenomenon corresponds to the vacuum alignment.
Additionally, since $v(\eta)$ breaks the extra flat directions,
masses of the quasi-NG modes that are  of the order of $\omega$ arise.
In this way, the quantum symmetry breaking of the 
$\text{U(1)}\times\text{SO(5)}$ symmetry indeed occurs.
 
\emph{Fate of topological defects.}---
Here, let us discuss the importance of the quasi-NG modes
in terms of topological defects.
If the symmetry breaking contribution \eqref{qsb} is negligible,
some topological defects predicted from $M$
become unstable.
This is because, under such a situation, 
topological defects are characterized by $\pi_n(\tilde{M})$ 
instead of $\pi_n(M)$, 
where $\pi_n$ represents the $n$th homotopy group.
To illustrate this in detail,
by using a modified Gross-Pitaevskii 
equation based on a local density approximation
\cite{fabrocini},
we examine the (in)stability of the $\mathbb{Z}_2$ vortex in 
the uniaxial nematic phase, 
which is
predicted from $\pi_1(\text{S}^2/\mathbb{Z}_2)=\mathbb{Z}_2$.
In the $\mathbb{Z}_2$ vortex, the spin-singlet trio amplitude
\beq
A_{30}&\equiv& 3\sqrt{6}(\phi_2\phi^2_{-1}+\phi^2_1\phi_{-2})/2
+\phi_0(\phi^2_0-3\phi_1\phi_{-1}-
6\phi_2\phi_{-2})\nonumber\\
&=&\cos3\eta
\label{trio}
\eeq
becomes zero
at the vortex core and has a finite value in the other region.
Initially,
the $\mathbb{Z}_2$ vortex is prepared under the condition that
the contribution \eqref{qsb}
cannot be neglected,
and, therefore, the vortex remains stable
and the quasi-NG modes are not  emitted spontaneously 
(Figs. \ref{fig1}(a) and (e)).
However, when the contribution  \eqref{qsb} is negligible,
the quasi-NG charge expands 
in a radial direction
and the vortex core structure disappears with time, which
indicates that the vortex decays due to the emission of the quasi-NG modes
(Figs. \ref{fig1}(b)-(d) and (f)-(h)).
We emphasize that
this drastic change in the vortex state 
is caused by quantum fluctuations.
\begin{figure}[t]
\includegraphics[width=3in]{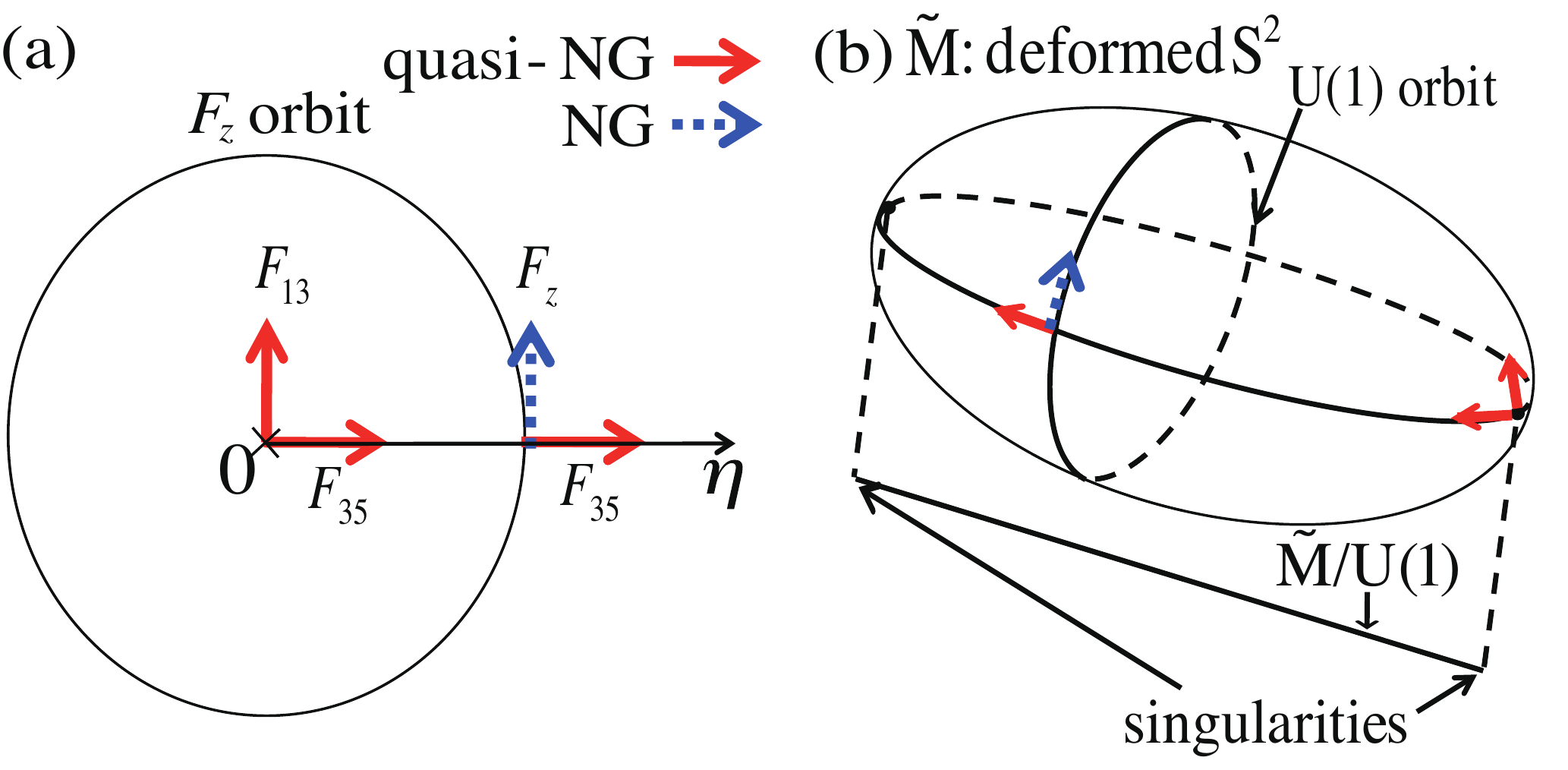}
\caption{(color online).
(a) Graphical interpretation of the NG--quasi-NG change within 
$0\le \eta <\pi/3$. Note that this is a vertical slice about
the directions along which the NG modes on
U(1), $F_{x}$, and $F_{y}$ fluctuate.
(b) A simple example of the NG-quasi-NG change in which
the target space is the rugby ball (homeomorphic to $\text{S}^2$),
the base space $\text{S}^2/\text{U(1)}$, and the structure group U(1).
Since U(1) symmetry is broken at the generic points, 
we have one NG mode along the U(1) orbit and one quasi-NG mode along
the base space $\text{S}^2/\text{U(1)}$ orthogonal to the U(1) orbit.
However, two quasi-NG modes appear at the singularities of the base space
where the U(1) symmetry is restored.
This is because neither of the two directions is generated by U(1) at
the U(1) fixed point.}
\label{fig2}
\end{figure} 

\emph{Geometrical interpretation.}---
The mathematical structure of the nematic phase
can be interpreted from the viewpoint of 
differential geometry \cite{nitta}.
The target space $\tilde{M}$ is parametrized by
both NG and quasi-NG modes.
To identify the quasi-NG modes, we eliminate the NG modes by
taking a quotient of $\tilde{M}$ by $G$.
Then, the quasi-NG mode corresponds to the orbit space
$\tilde{M}/G$
and it is parametrized by quantities made of $\psi_m$,
which are $G$-invariant but not $\tilde{G}$-invariant. 
There should exist $n$  such invariants with $n$ defined in 
\eqref{number-png} at generic points, where the dimension of unbroken
symmetry dim($H$)
is the lowest.
For the nematic phase, where the generic points are the
dihedral-2 and biaxial nematic phases with $n=1$,
there should be 
one polynomial to describe $\tilde{M}/G$. 
In fact, we can show that this is uniquely determined to be
the absolute value of the 
spin-singlet trio amplitude $|A_{30}|$. 
From this result, we find that the orbit space $\tilde{M}/G$
can be characterized by $\eta$.
Physically, the direction of $\eta$ is that of the quasi-NG mode, 
which is generated by $F_{35}$.
The target space $\tilde{M}$ can be regarded as a
(singular) fiber bundle over the base space $\tilde{M}/G$,
with fiber $M$ and structure group $G$.
The fiber shrinks at certain points where $H$ is 
enhanced and which correspond to the singular points in 
$\tilde{M}/G$.
Since the number of NG modes is reduced at such points,
some of the NG modes need to change to quasi-NG modes
to keep the total number of soft modes unchanged.
In fact, the NG mode generated by $F_z$ changes to
the quasi-NG mode generated by $F_{13}$ in the uniaxial nematic phase,
which corresponds to a singularity in $\tilde{M}/G$,
as shown in Fig. \ref{fig2}.

In conclusion, we have shown that quasi-NG modes can be
realized with the spin-2 nematic phase because
the vacuum manifold at the zeroth order is enlarged to $\tilde{M}$ and
each $M$ is a submanifold of $\tilde{M}$.
This corresponds to the vacuum alignment problem, where a unique vacuum
is selected by considering quantum fluctuations.
We expect that 
the phenomena illustrated in Fig. \ref{fig1} open up a new possibility in quantum
vortices that can 
be tested experimentally by controlling quantum fluctuations.

This work was supported by KAKENHI (22340114, 20740141, and 22103005),
Global COE Program ``the Physical Sciences Frontier'' and
the Photon Frontier Network Program, MEXT, Japan.

 

\begin{thebibliography}{99}
\bibitem{note1}
Although this was originally  called the pseudo-NG mode,
this term is currently used to represent a soft mode which
has a small mass attributed to explicitly symmetry breaking terms
at the zeroth order. 
Hence, we refer to it as the quasi-NG mode
to avoid confusion with the pseudo-NG mode.
\end{thebibliography}
\end{document}